\documentclass[conference]{IEEEtran}
\IEEEoverridecommandlockouts

\usepackage[linesnumbered,ruled,vlined]{algorithm2e}

\usepackage{cite}
\usepackage{amsmath,amssymb,amsfonts}
\usepackage{algorithmic}
\usepackage{graphicx}
\usepackage{textcomp}
\usepackage{xcolor}
\usepackage{subfig}
\usepackage{booktabs}
\usepackage{multirow,multicol}
\usepackage{hyperref}
\def\BibTeX{{\rm B\kern-.05em{\sc i\kern-.025em b}\kern-.08em
    T\kern-.1667em\lower.7ex\hbox{E}\kern-.125emX}}
\begin{document}

\title{TurboBias 2.0: Streaming Context-Biasing \\ for Production-Efficient ASR Systems}

\author{
\IEEEauthorblockN{
    Vladimir Bataev\IEEEauthorrefmark{1}\IEEEauthorrefmark{3}\IEEEauthorrefmark{4}, 
    Lilit Grigoryan\IEEEauthorrefmark{1}\IEEEauthorrefmark{3},
    Andrei Andrusenko\IEEEauthorrefmark{1}\IEEEauthorrefmark{3},
    Nikolay Karpov\IEEEauthorrefmark{1}, 
    Vitaly Lavrukhin\IEEEauthorrefmark{2},
    Boris Ginsburg\IEEEauthorrefmark{2}}
\newline
\IEEEauthorblockN{
    \IEEEauthorrefmark{1}\textit{NVIDIA}, Yerevan, Armenia
    \IEEEauthorrefmark{2}\textit{NVIDIA}, Santa Clara, USA
}
\IEEEauthorblockN{
    \IEEEauthorrefmark{3}Equal contribution
    \IEEEauthorrefmark{4}Corresponding author, \textit{vbataev@nvidia.com}
}
}

\maketitle

\begin{abstract}
Contextualization is essential for production automatic speech recognition (ASR) systems, where user-provided phrases must be recognized accurately under strict latency constraints. Although many context-biasing methods improve recognition accuracy, they often do not address the practical requirements of modern production ASR systems: streaming inference, efficient batched decoding, user-specific context lists, and low runtime overhead. We propose TurboBias 2.0, a production-oriented framework for efficient phrase boosting in Transducer-based ASR systems. The framework extends GPU-accelerated TurboBias with a case-insensitive boosting graph and per-stream batched decoding, allowing each utterance in a batch to use an independent context-biasing configuration. This enables personalized context biasing for multiple simultaneous users without sharing or mixing their context lists. The proposed framework supports both offline and streaming inference and can be used with greedy and beam-search decoding. Experiments show that TurboBias 2.0 improves contextual phrase recognition while preserving low latency and high throughput.
\end{abstract}


\section{Introduction}

Modern automatic speech recognition (ASR) systems achieve strong accuracy on general-purpose benchmarks, but they still often struggle with rare, domain-specific, and user-defined terminology. These terms may be infrequent in training data, yet they are often crucial for transcript usability. Therefore, context biasing remains an important component of production ASR systems, especially for meetings, earnings calls, medical transcription, and voice assistants.

A wide range of context-biasing approaches has been proposed. Deep-fusion methods inject contextual information directly into the ASR model through additional attention mechanisms, adapters, or specialized training objectives \cite{Pundak2018DeepCE,Yang2023PromptASRFC,Jain2020ContextualRF,Le2021ContextualizedSE,Harding2023SelectiveBW}. These methods can be accurate but often require model modifications and retraining. Recent speech and speech-language models can also use textual prompts with contextual phrases \cite{Wang2023SLMBT,Chen2023SALMSL}, but their behavior may be sensitive to prompt design, context-list length, and latency constraints. In contrast, shallow-fusion and keyword-boosting methods modify decoding scores for hypotheses that match a predefined context list \cite{Zhao2019ShallowFusionEC,Jung2021SpellMN,Huang2024ImprovingNB,He2018StreamingES}. This makes them attractive for production because they can often be applied without retraining the ASR model.

Several recent works have focused on efficient contextual ASR. CTC-based Word Spotter (CTC-WS) detects keywords by matching CTC log-probabilities against a context graph and then correcting corresponding spans in the greedy ASR output \cite{Andrusenko2024ctcws}. While effective, this approach requires a CTC scoring path and a post-recognition correction stage. For Transducer-based ASR (RNNT~\cite{graves2012rnnt}, TDT~\cite{Xu2023EfficientST}), it is most natural for hybrid Transducer-CTC models or pipelines with a separate CTC model. TurboBias introduced a GPU-accelerated phrase-boosting tree for shallow-fusion context biasing and demonstrated efficient support for CTC, Transducer, and Attention Encoder-Decoder models \cite{Andrusenko2025TurboBias}. More recently, Contextual Earnings-22 introduced a public contextual ASR benchmark with realistic custom vocabulary from earnings calls and evaluation settings with both local context and global context lists containing distractors \cite{Durmus2026ContextualEA}. Despite this progress, several requirements of high-load streaming production ASR remain insufficiently addressed.

First, modern ASR systems often produce formatted transcripts with punctuation and capitalization, so context biasing should be robust to casing differences between user-provided phrases and model output. Expanding each phrase with lowercase, uppercase, and title-case variants is possible, but increases tree size, compilation time, and memory usage. A more practical solution is to support case-insensitive matching directly in the boosting structure.

Second, production workloads often require personalized context biasing for many simultaneous users. In batched ASR, different streams may correspond to different meetings, medical conversations, earnings calls, or customer-support sessions, each with its own context list. A shared boosting tree can introduce distractors and false acceptances across streams, so a production-ready system should support independent per-stream boosting configurations within a single GPU batch.

Third, many applications require streaming recognition with strict latency constraints. Context biasing should therefore work in streaming greedy and beam-search decoding, not only offline beam search. This is especially important for Transducer-based models, which are widely used in production streaming ASR. The additional overhead from context biasing must also remain small relative to overall ASR inference time.

In this work, we propose TurboBias 2.0, a production-oriented extension of TurboBias framework for efficient context biasing in streaming Transducer ASR. The proposed framework introduces a case-insensitive GPU phrase-boosting graph, supports phrase boosting in streaming greedy and beam-search decoding, and enables per-stream boosting graphs for batched inference. This allows multiple simultaneous streams to use independent context-biasing configurations while preserving efficient GPU execution and low-latency streaming recognition.
We evaluate TurboBias 2.0 on Contextual Earnings-22 and an internal medical-domain test set using modern streaming ASR models. We compare the proposed case-insensitive graph with case-sensitive and phrase-expanded alternatives, analyze offline and streaming decoding modes, and study the trade-off between keyword F-score, beam size, and streaming latency.

The main contribution of the paper is:
\begin{itemize}
    \item Case-insensitive GPU phrase-boosting graph that avoids an additional phrase-list expansion such as lowercase, uppercase, and title-case variants.
    \item Per-stream context-biasing graphs in batched inference, where each user/request has an independent phrase list for efficient batched decoding.
    \item Streaming Transducer beam-search support to improve boosting performance in real-time speech applications.
\end{itemize}

The proposed TurboBias 2.0 framework is open-sourced as part of the NVIDIA NeMo toolkit~\cite{kuchaiev2019nemo}\footnote{\scriptsize
NeMo PRs: 
\href{https://github.com/NVIDIA-NeMo/Speech/pull/15800}{\#15800},
\href{https://github.com/NVIDIA-NeMo/Speech/pull/15125}{\#15125}, and
\href{https://github.com/NVIDIA-NeMo/Speech/pull/15753}{\#15753}.}.


\section{Method description}

\subsection{Case-Insensitive Boosting}

Original TurboBias represents a context list as a GPU phrase-boosting tree with suffix links, where each phrase is tokenized by the ASR model tokenizer and every matched token receives a shallow-fusion bonus during decoding \cite{Andrusenko2025TurboBias}. This construction is efficient, but it is case-sensitive: a boosting entry \textit{hello} does not necessarily match hypotheses such as \textit{Hello} or \textit{HELLO}. This is a practical limitation for modern formatted ASR systems, where the model may emit capitalization at sentence boundaries or for named entities, while user-provided context phrases are often unformatted.

\begin{figure}[t]
    \centering
    \vspace{-10pt}
    \subfloat[\centering Character tokenization]{{
        \includegraphics[width=0.8\linewidth]{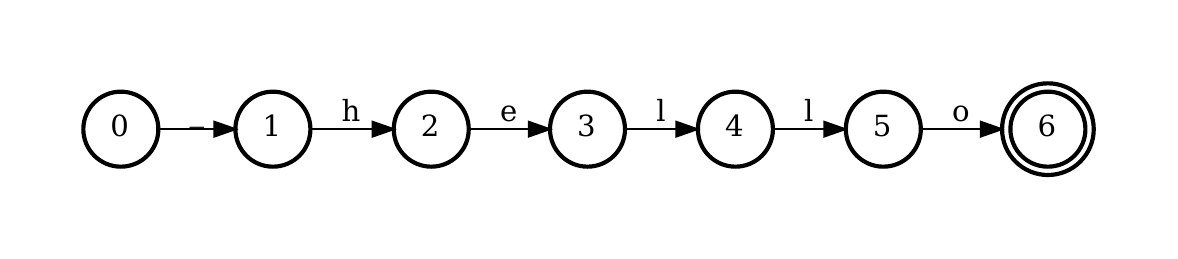}
        \label{fig:tokenization_char}
    }}
    \qquad
    \subfloat[\centering BPE tokenization (greedy)]{{
        \includegraphics[width=0.5\linewidth]{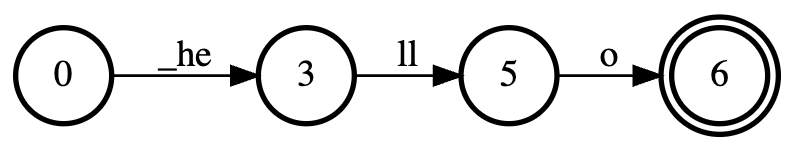}
        \label{fig:tokenization_bpe_greedy}
    }}
    \qquad
    \centering
    \vspace{-10pt}
    \subfloat[\centering Variative BPE tokenization]{{
        \includegraphics[width=0.8\linewidth]{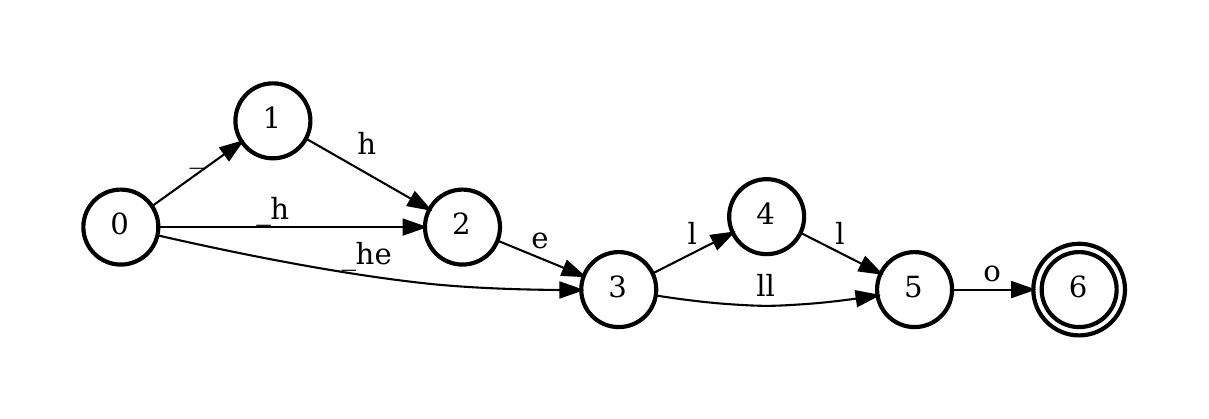}
        \label{fig:tokenization_var_bpe}
    }}
    \qquad
    \centering
    \vspace{-10pt}
    \subfloat[\centering Case-insensitive character tokenization]{{
        \includegraphics[width=0.8\linewidth]{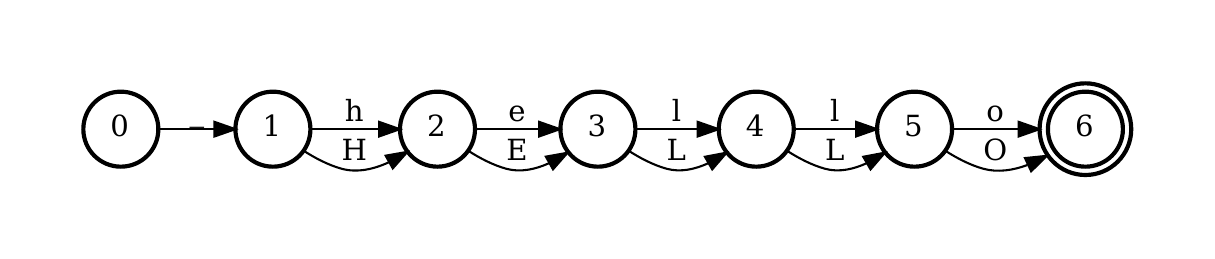}
        \label{fig:tokenization_char_ci}
    }}
    \qquad
    \centering
    \subfloat[\centering Case-Insensitive variative BPE tokenization]{{
        \includegraphics[width=0.8\linewidth]{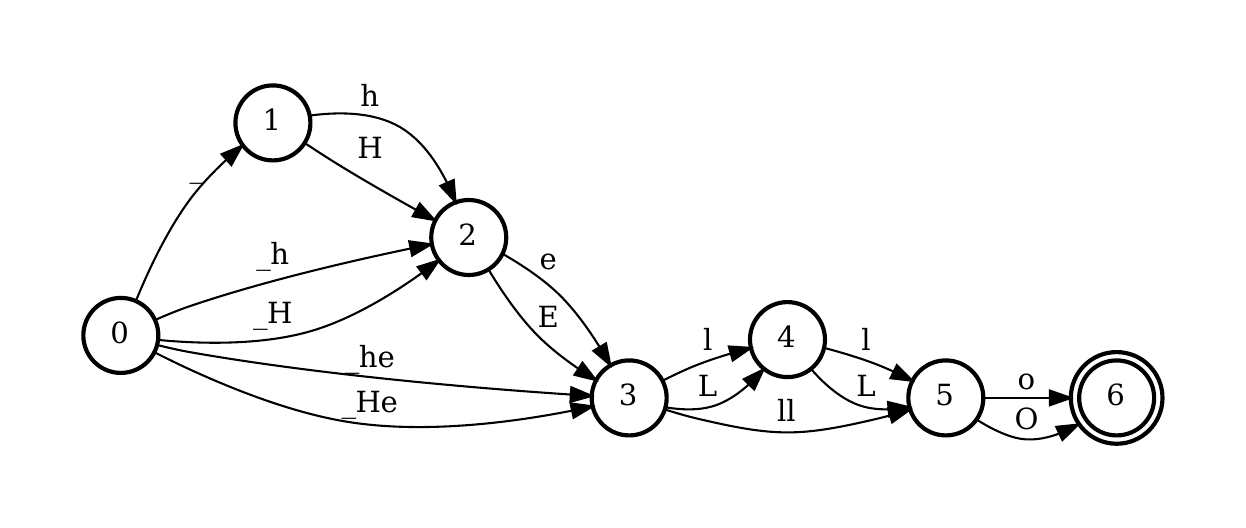}
        \label{fig:tokenization_ci_bpe}
    }}
    \vspace{-5pt}
    \caption{Tokenization graphs for the word \textit{hello}.}
    \label{fig:ci_tokenization}
    \vspace{-10pt}
\end{figure}

A direct workaround is to insert multiple spellings of every phrase, for example lowercase, title-case, and uppercase variants. However, this increases the number of paths in the boosting tree and still does not cover all mixed-case outputs. For character-tokenized models, a cleaner construction is straightforward: each character arc can be replaced with a set of two parallel arcs for case variants, as illustrated in Fig.~\ref{fig:tokenization_char_ci}. This preserves a compact graph and makes the match independent of the emitted capitalization.

The same idea cannot be applied directly to greedy BPE tokenizations. BPE vocabularies do not generally contain a case variant for every token; a token such as \textit{\_he} may exist while \textit{\_He} or \textit{\_HE} may not, and the tokenizer may segment the cased spelling differently. We therefore first convert each phrase into a variative BPE representation, as illustrated in Fig.~\ref{fig:tokenization_var_bpe}. Given a phrase, we take its greedy SentencePiece tokenization~\cite{sennrich2016bpe,Kudo2018SentencePiece} and decompose each token into its canonical character-level pieces. For every position in this character chain, we add the corresponding single-piece token arcs, and for every vocabulary token whose canonical decomposition spans a contiguous segment of the chain, we add a merged arc over that segment. Since SentencePiece tokenizers are built from character pieces, the character-level chain is available for the models considered here, and vocabulary lookup is sufficient to enumerate valid merged arcs.

For case-insensitive boosting, the canonical representation is lowercased. Each vocabulary token is mapped to the canonical sequence obtained by lowercasing its piece before decomposition, and it is allowed as an arc whenever this canonical sequence matches the lowercased phrase path. The graph therefore accepts different cased BPE segmentations that correspond to the same lowercased phrase, as shown in Fig.~\ref{fig:tokenization_ci_bpe}. The resulting structure is no longer a tree, because several arcs may skip over different numbers of character positions and converge to the same state. It remains acyclic apart from the explicit suffix links, so it can still be compiled into the same NGPU-LM-based data structure~\cite{article:ngpulm} used by TurboBias.

The variative representation also changes how boosting scores should be assigned. In the original tree, an arc corresponds to one tokenizer token from the greedy phrase representation, so the TurboBias token score can be stored directly on that arc. In the case-insensitive variative graph, paths may contain different numbers of arcs for the same phrase span. Assigning the same score to every arc would make character-by-character paths receive a larger total bonus than merged-token paths, while assigning score only to merged arcs would delay the bonus and weaken partial matches.

We address this by storing a cumulative score potential on states and defining each arc score as the difference between the destination and source potentials. The potentials are computed from the lowercased greedy BPE tokenization, which serves as the scoring backbone. If the $k$-th greedy BPE token has TurboBias score $w_k$ and decomposes into $m_k$ character pieces, we distribute $w_k$ over the internal character states using
\begin{equation}
\Delta_{k,i} =
w_k
\frac{
\exp((i + 1)^{\tau})
}{
\sum_{j=0}^{m_k - 1} \exp((j + 1)^{\tau})
},
\quad i = 0,\ldots,m_k-1 ,
\end{equation}
where $\tau$ controls how concentrated the score is near the end of the token. Setting $\tau=0$ gives a uniform split, while larger values approach the original TurboBias behavior where most of the score is applied when the full BPE token has been matched. In practice, the endpoint-concentrated setting is conservative for greedy decoding, while flatter distributions can help beam search because partial phrase matches receive boosting earlier. Since all cased and merged alternatives ending in the same state share the same potential, every accepted spelling receives the same total phrase bonus, independent of its BPE segmentation.

\subsection{Per-Stream Context Biasing}

In production ASR services, context phrases are often request-specific. A single server may batch streams from multiple customers, where each stream has a small list of names, product terms, medical entities, or other phrases that should be boosted only for that stream. This creates two requirements not addressed by a single global boosting tree. First, a phrase list from one customer must not affect recognition quality for other streams in the same batch. Second, phrase lists must be added and removed dynamically as streams start and finish, without reloading the ASR server or reinstantiating the decoder.

TurboBias 2.0 addresses this with a per-stream context-biasing model built on the same NGPU-LM/TurboBias tensor representation. Each boosting tree is still compiled as a weighted acceptor with arc-related tensors, such as labels, target states, and arc weights, and state-related tensors, such as outgoing-arc ranges, suffix-link targets, suffix-link weights, and final weights. Instead of storing one active tree, we concatenate the tensors of all active boosting trees into merged multi-model storage. For each model id, the multi-model stores the corresponding state and arc ranges as offsets into the merged tensors, together with the boosting weight for that model.

During batched decoding, every stream carries a biasing model id. A non-negative id selects the phrase list for that stream, while id $-1$ denotes no biasing. The fused scoring operation receives the current TurboBias states and the per-stream model ids. For each batch item, the GPU kernel first reads the state and arc offsets associated with its model id, shifts the tensor pointers to the selected range, and then runs the same suffix-link traversal used by the single-tree TurboBias advance operation. If the model id is $-1$, the kernel returns zero boosting scores and no valid next states, so the stream is decoded exactly as an unbiased stream. Thus, different streams in the same GPU batch can use independent phrase lists without sharing distractors.

The merged representation supports frequent model lifecycle updates. Adding a model appends its arc and state tensors to the merged storage and records its offsets. Removing a model marks its id inactive and shifts later arc and state ranges left to fill the removed interval; since these are contiguous tensor operations, the update is efficiently parallelized on GPU. Model ids are tracked separately from physical tensor offsets, allowing the decoder to keep a stable per-stream id interface while the underlying storage is compacted.

This layout preserves the computational profile of a single global boosting tree. Once the merged storage is on GPU, each stream still performs one boosting advance over one selected tree and writes one score vector over the vocabulary. Compared with the single-tree case, the additional work is limited to reading the per-stream model id and the two offsets before the standard arc/state traversal. In our measurements this does not introduce observable decoding-speed degradation relative to using one global tree, while avoiding cross-stream quality regressions from irrelevant phrases.

The main remaining cost is compiling and transferring boosting models when streams are created. In realistic deployments the per-stream lists are usually small, and an individual boosting model is typically below 1 MB, so keeping hundreds of compiled models is practical. To reduce model-lifecycle overhead, we use a three-level cache strategy: disk storage for persistent precompiled boosting trees, CPU-memory storage for compiled models that can be copied into the merged GPU model on demand, and decoder-level registration for models that are known before a recognition session. The first level is useful for large or repeated phrase lists, the memory level keeps dynamic per-stream updates within a small overhead, and decoder-level registration fully removes the update cost from the decoding loop.

\subsection{Streaming Beam Search}

\cite{Grigoryan2025Pushing} introduced offline batched decoding algorithms for Transducer-based models, narrowing the latency gap between greedy and beam-search decoding through tree-structured hypothesis storage, fully GPU-based computation, and CUDA Graph execution. We extend this approach to real-time streaming ASR. In contrast to offline decoding, streaming decoding processes encoder outputs chunk by chunk, requiring the beam-search state to be maintained across chunk boundaries. Our streaming extension preserves this continuity by storing finalized chunk-level hypotheses in a flattened tree together with their accumulated scores and boundary decoder states. These stored states are then used as anchors for continuing, merging, and pruning newly expanded hypotheses in subsequent chunks. The implementation remains compatible with CUDA Graphs and fully GPU-based batched operations, enabling low-latency streaming Transducer beam search while retaining the execution pattern of the offline algorithm.

\section{Experimental setup}

\subsection{ASR models}

For evaluation, we use publicly available English Transducer-based ASR models. As a strong offline baseline, we used popular  \texttt{parakeet-tdt-0.6b-v2}~\cite{model:parakeet-tdt-0.6b-v2} (TDT-v2), a 600M-parameter model based on a FastConformer encoder~\cite{rekesh2023fastconformer} and a TDT decoder~\cite{Xu2023EfficientST}. 

For low-latency streaming experiments, we use \texttt{nemotron-speech-streaming-en-0.6b}~\cite{model:nemotron-speech-streaming-en-0.6b} (Nemotron-streaming), a 600M-parameter model with a cache-aware FastConformer encoder and an RNNT decoder. The model processes audio chunks while reusing cached encoder states from previous chunks, making it a strong streaming baseline for the low-latency end of the accuracy--latency trade-off.

We also evaluate \texttt{parakeet-unified-en-0.6b}~\cite{model:parakeet-unified-en-0.6b} (Unified), a 600M-parameter model supporting both offline and streaming inference. The model uses a FastConformer encoder and an RNNT decoder, with parameters shared between offline and streaming modes. In streaming mode, it applies chunk-limited self-attention and dynamic chunked convolutions, while MCR-RNNT training reduces the gap between offline and streaming recognition~\cite{Andrusenko2026ReducingTO}.

All models support punctuation and capitalization, which is important for studying case-insensitive phrase boosting. 

\subsection{Evaluation data}

We evaluate the proposed context-biasing framework on two English datasets representing different production-oriented scenarios. The first dataset is Contextual Earnings-22 (referred to as C-Earnings22 in the rest of the paper), a public benchmark built on Earnings-22 for evaluating contextual ASR with a custom vocabulary \cite{Durmus2026ContextualEA}. It focuses on earnings-call transcription, where accurate recognition of person, company, and product names is critical. Each sample is a 15-second clip centered around at least one contextual keyword, with a manually reviewed transcript and context list. The benchmark provides two context regimes: a local list containing only keywords in the target clip, and a global list built from the full call-level keyword inventory with realistic distractors.

We also use an internal medical-domain test set to evaluate the method on specialized terminology. The dataset contains approximately 3 hours of English medical speech with contextual phrases such as diseases, procedures and medications clinical conditions. 

Table~\ref{tab:dataset_stats} summarizes the evaluation datasets used in this work. The validation set is used for selecting context-biasing parameters, while the test set is used for final evaluation.

\subsection{Compared methods}

In addition to the proposed TurboBias 2.0 variants described in the previous section, we evaluate a CTC-WS-based baseline\footnote{\scriptsize{\url{https://github.com/NVIDIA-NeMo/Speech/blob/main/scripts/asr_context_biasing/eval_greedy_decoding_with_context_biasing.py}}} using the setup from the Contextual Earnings-22 benchmark \cite{Durmus2026ContextualEA}. The main ASR output is produced by \texttt{parakeet-tdt-0.6b-v2}~\cite{model:parakeet-tdt-0.6b-v2} using greedy TDT decoding, while keyword spotting is performed with a separate \texttt{parakeet-tdt-ctc-110m}~\cite{model:parakeet-tdt-ctc-110m} (TDT-CTC) model. 
This baseline uses two independent ASR inference paths: one for the initial TDT transcript and one for CTC-based contextual phrase detection for the final text correction.

The Contextual Earnings-22 paper reports a modified CTC-WS integration strategy, but does not provide any implementation details required for an exact reproduction. Therefore, we use the same publicly available ASR models and the standard CTC-WS method as closely as possible.

\begin{table}[t!]
  \centering
  \caption{Statistics of evaluation datasets (test part).}
  \label{tab:dataset_stats}
  \resizebox{\columnwidth}{!}{%
  \begin{tabular}{lcc}
    \toprule
    \textbf{Metric} & \textbf{C-Earnings22} & \textbf{Internal Medical} \\
    \midrule
    Domain           & Earnings & Medical \\
    Size (hours)     & 2.63     & 3.05    \\
    Keyword inst.    & 1{,}259  & 2{,}630 \\
    Unique keywords  & 738      & 489     \\
    \midrule
    \multirow{2}{*}{Example keywords}
       & \textit{Husqvarna, NASDAQ,}  & \textit{linezolid, quinapril,}  \\
       & \textit{GlobalSign} & \textit{abciximab}  \\
    \bottomrule
  \end{tabular}
  }
\vspace{-10pt}
\end{table}

\begin{figure*}[t]
    \centering 
    \includegraphics[width=1.0\textwidth]{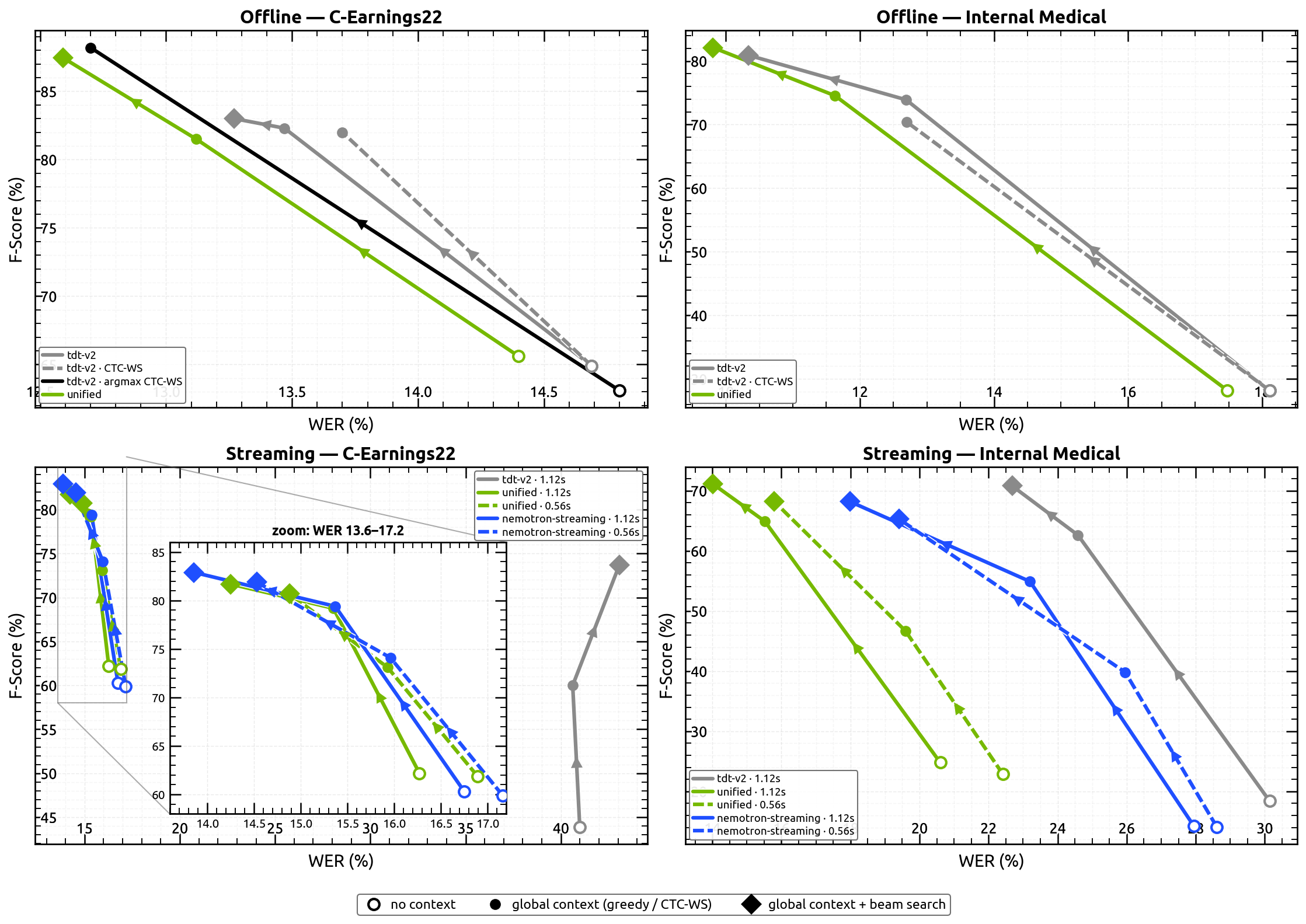}
    \caption{Contextual Biasing: WER vs F-Score}
    \label{fig:main-plot}
    \vspace{-10pt}
\end{figure*}

\subsection{Metrics}

We evaluate recognition quality using WER on the full ASR output and keyword precision, recall, and F-score on contextual phrases only. A phrase is counted as correctly recognized if it appears in the hypothesis and matches the corresponding reference phrase after text normalization. Let $N_{\mathrm{TP}}$, $N_{\mathrm{FP}}$, and $N_{\mathrm{FN}}$ denote the number of true-positive, false-positive, and false-negative contextual phrase matches. Then,
\begin{equation}
P = \frac{N_{\mathrm{TP}}}{N_{\mathrm{TP}} + N_{\mathrm{FP}}},
\quad
R = \frac{N_{\mathrm{TP}}}{N_{\mathrm{TP}} + N_{\mathrm{FN}}},
\quad
F = \frac{2PR}{P + R}.
\end{equation}

Decoding efficiency is measured with inverse real-time factor:
\begin{equation}
\mathrm{RTFx} = \frac{T_{\mathrm{audio}}}{T_{\mathrm{eval}}},
\end{equation}
where $T_{\mathrm{audio}}$ is the total audio duration and $T_{\mathrm{eval}}$ is the ASR decoding time. Higher RTFx indicates faster decoding. We use a single NVIDIA RTX A5000 24GB GPU. 

We evaluate three context-list configurations. In the \textit{full} setting, a single context list is built from all contextual phrases for all streams. In the \textit{per-stream global} setting, each stream has its own context list containing only the phrases associated with that stream, simulating personalized high-load production inference. In the \textit{per-stream local} setting, the context list is further restricted to phrases relevant to the current audio segment within each stream.

For streaming experiments, we use either chunk-based (TDT-v2 and Unified) or cache-aware decoding (Nemotron-streaming). We report streaming latency as the worst-case algorithmic latency of the ASR model, computed as the sum of the chunk size and the right context when used.

All models are evaluated with both greedy
(label-looping~\cite{bataev2024labellooping,galvez24_speedoflight})
and beam-search (ASLD++~\cite{alsd,Grigoryan2025Pushing}, with $beam\_size=32$) decoding. Validation sets are used to select boosting parameters, and final results are reported on the test sets.

\section{Results}

Figure~\ref{fig:main-plot} summarizes the main comparison between CTC-WS-based contextual biasing and direct Transducer phrase boosting through TurboBias 2.0. We include the Argmax CTC-WS result from the Contextual Earnings-22 benchmark as a reference point. This system is not a single-model contextual decoder: it combines a TDT-v2 ASR transcript with keyword detections from a CTC model trained jointly with a TDT objective. The benchmark paper also describes a modified CTC-WS integration, but the implementation details and scoring scripts needed for exact reproduction are unavailable. Therefore, we report the published Argmax result separately and compare it with our closest reproduction using TDT-v2 with CTC-WS. This reproduced setup improves the no-context TDT-v2 baseline on Earnings22 from 64.9 to 82.0 F-score and reduces WER from 14.7\% to 13.7\%, but remains below the reported Argmax point of 88.2 F-score and 12.7\% WER. The result confirms the effectiveness of CTC-WS, while also highlighting its practical complexity: it requires an auxiliary CTC inference path, a suitable hybrid TDT-CTC model, and a post-recognition text-correction stage.

TurboBias 2.0 improves contextual recognition by applying the biasing scores directly inside Transducer decoding. For the offline TDT-v2 model, per-stream global boosting gives substantial gains over the unadapted model, reaching 83.0 F-score and 13.3\% WER on Earnings22 with beam search. The strongest overall results are obtained with the Unified ASR model. In the offline setting, the Unified model reaches 87.5 F-score and 12.6\% WER on Earnings22, which is close to the reported Argmax CTC-WS F-score and gives a lower WER. It also substantially outperforms our TDT-v2 + CTC-WS reproduction. On the internal medical set, the same trend holds: direct TurboBias 2.0 boosting produces large improvements in contextual F-score while also reducing WER, showing that the method transfers to a domain with specialized terminology.

The streaming results in Fig.~\ref{fig:main-plot} show that the choice of ASR model is critical under latency constraints. The streaming TDT-v2 configuration suffers a large quality degradation relative to its offline mode, and context biasing cannot fully compensate for this loss. In contrast, the Unified model preserves much stronger streaming accuracy, and the relative gains from TurboBias 2.0 are comparable to those observed offline. On Earnings22, the Unified streaming model at 1.12s worst-case latency (chunk 0.56s, right context 0.56s) improves from 62.2 to 81.7 F-score and from 16.3\% to 14.3\% WER after global boosting and beam search. The lower-latency 0.56s (chunk 0.16s, right context 0.40s) configuration follows the same trend. The Nemotron-streaming model performs similarly on Earnings22 and obtains a slightly larger benefit from beam search. However, on the internal medical set, the Unified model is substantially stronger: at 1.12s latency it reaches 71.1 F-score and 14.0\% WER, compared with 68.2 F-score and 18.0\% WER for Nemotron-streaming.

\begin{table}[t]
\centering
\small
\caption{Case-sensitive vs. case-insensitive boosting under different biasing scenarios on Earnings22 using \texttt{parakeet-unified-en-0.6b}.}
\label{tab:boosting_ablation}
\resizebox{\columnwidth}{!}{
\begin{tabular}{lllccc}
\toprule
\textbf{Decoding} & \textbf{Context} & \textbf{Case} & \textbf{Per-str} & \textbf{F-Score (P/R)$\uparrow$} & \textbf{WER$\downarrow$} \\
\midrule
Greedy & -- & -- & -- & 65.2 (92.6/50.3) & 14.4 \\
Beam   & -- & -- & -- & 65.9 (92.7/51.2) & 14.0 \\
\midrule
\multirow{3}{*}{Greedy}
& \multirow{3}{*}{Full}
& lower                & -- & 66.0 (91.7/51.5) & 14.4 \\
& & target             & -- & 74.8 (89.0/64.5) & 13.7 \\
& & \underline{case-ins.} & -- & 75.4 (89.9/64.8) & 13.7 \\
\cmidrule(lr){2-6}
\multirow{3}{*}{Beam}
& \multirow{3}{*}{Full}
& lower                & -- & 72.5 (90.9/60.3) & 13.3 \\
& & target             & -- & 80.6 (88.0/74.3) & 12.8 \\
& & \underline{case-ins.} & -- & 81.4 (89.2/74.8) & 12.8 \\
\midrule
\multirow{2}{*}{Greedy}
& Global & \underline{case-ins.} & $\checkmark$ & 81.5 (91.5/73.4) & 13.1 \\
& Local  & \underline{case-ins.} & $\checkmark$ & 85.4 (96.6/76.6) & 12.7 \\
\cmidrule(lr){2-6}
\multirow{2}{*}{Beam}
& Global & \underline{case-ins.} & $\checkmark$ & 87.5 (92.4/83.0) & 12.6 \\
& Local  & \underline{case-ins.} & $\checkmark$ & 91.2 (95.8/87.0) & 12.2 \\
\bottomrule
\end{tabular}
}
\vspace{-10pt}
\end{table}

Table~\ref{tab:boosting_ablation} isolates the contribution of case-insensitive boosting and per-stream boosting. With the original case-sensitive boosting tree, performance depends strongly on the spelling of the input context list. Lowercase phrases give only a small improvement over the no-context baseline in greedy decoding, whereas target-cased phrases, such as capitalized proper nouns, increase F-score from 65.2 to 74.8 in greedy decoding and from 65.9 to 80.6 with beam search. The proposed case-insensitive tree removes this dependence on phrase-list capitalization and reaches 75.4 and 81.4 F-score in greedy and beam-search decoding, respectively. Thus, it slightly surpasses the target-cased case-sensitive tree while remaining robust when the provided context list is lowercase or otherwise incorrectly cased or contains incomplete cased variants.

Per-stream boosting further improves recognition by avoiding irrelevant phrases from other streams in the same batch. With per-stream global context, where each stream receives its target phrases together with realistic distractors, F-score increases to 81.5 in greedy decoding and 87.5 with beam search, while WER decreases to 13.1\% and 12.6\%, respectively. The per-stream local setting, which contains only target phrases for each utterance, gives the best absolute numbers: 85.4 F-score with greedy decoding and 91.2 with beam search. However, this setting assumes knowledge of the exact phrases appearing in the utterance and is therefore less representative of most production deployments. We therefore treat per-stream global boosting as the main realistic operating point.

\begin{table}[t]
\centering
\small
\caption{TurboBias 2.0 decoding speed on Earnings22 using \texttt{parakeet-unified-en-0.6b} with 128 streams.}
\label{tab:boosting_speed}
\setlength{\tabcolsep}{3.5pt}
\begin{tabular}{lcccc}
\toprule
\textbf{Biasing} & \textbf{Cache} & \textbf{Per-stream} & \textbf{F-score$\uparrow$} & \textbf{RTFx$\uparrow$} \\
\midrule
\multicolumn{5}{l}{\textbf{Greedy decoding}} \\
\midrule
No biasing & --      & --           & 65.2 & 1812 \\
Full       & --      & --           & 75.4 & 1726 \\
\cmidrule(lr){1-5}
Full       & --      & $\checkmark$ & 75.4 & 61 \\
Full       & disk    & $\checkmark$ & 75.4 & 794 \\
Full       & memory  & $\checkmark$ & 75.4 & 1493 \\
Full       & decoder & $\checkmark$ & 75.4 & 1717 \\
\cmidrule(lr){1-5}
Global     & --      & $\checkmark$ & 81.5 & 695 \\
Global     & disk    & $\checkmark$ & 81.5 & 851 \\
Global     & memory  & $\checkmark$ & 81.5 & 1467 \\
Global     & decoder & $\checkmark$ & 81.5 & 1714 \\
\midrule
\multicolumn{5}{l}{\textbf{Beam-search decoding}} \\
\midrule
No biasing & --      & --           & 65.9 & 1150 \\
Full       & --      & --           & 81.4 & 1005 \\
\cmidrule(lr){1-5}
Full       & --      & $\checkmark$ & 81.4 & 59 \\
Full       & disk    & $\checkmark$ & 81.4 & 581 \\
Full       & memory  & $\checkmark$ & 81.4 & 905 \\
Full       & decoder & $\checkmark$ & 81.4 & 989 \\
\cmidrule(lr){1-5}
Global     & --      & $\checkmark$ & 87.5 & 534 \\
Global     & disk    & $\checkmark$ & 87.5 & 631 \\
Global     & memory  & $\checkmark$ & 87.5 & 889 \\
Global     & decoder & $\checkmark$ & 87.5 & 988 \\
\bottomrule
\end{tabular}
\vspace{-10pt}
\end{table}

Table~\ref{tab:boosting_speed} summarizes the speed evaluation for 128 concurrent streams. The qualitative behavior is consistent across decoding modes. Disk caching can reduce setup time for repeated or large context lists, but its benefit depends on storage throughput and is not always observable for small per-stream lists. Keeping compiled models in CPU memory introduces only a small recognition-speed overhead of approximately 9-17\%. When context models are registered at the decoder level before recognition, the model-loading and update overhead is fully hidden, and decoding speed remains comparable to the global-boosting baselines.

\section{Conclusion}

We presented TurboBias 2.0, an efficient framework for contextual phrase boosting in Transducer-based ASR systems. The proposed case-insensitive boosting graph makes decoding robust to capitalization differences between user phrase lists and formatted ASR output, without expanding phrases into multiple case variants. We also introduced per-stream context biasing for batched inference, allowing each stream to use an independent context list and avoiding interference from unrelated distractors.
We further extended phrase boosting to streaming inference and adapted GPU beam search for streaming Transducer decoding. Experiments on Earnings22 and an internal medical-domain set show consistent improvements in contextual F-score and WER, with beam search providing substantial gains over greedy decoding in both offline and streaming settings. These results indicate that decoder-level phrase boosting can deliver strong contextual accuracy with low overhead, without auxiliary CTC inference, post-recognition correction, or model retraining.

\clearpage
\bibliographystyle{IEEEtran}
\bibliography{refs}

\end{document}